\documentclass[
  aps,
  prb,
  twocolumn,
  superscriptaddress,
  nofootinbib,
  floatfix
]{revtex4-2}

\usepackage{graphicx}
\usepackage{amsmath}
\usepackage{amssymb}
\usepackage{bm}
\usepackage{xcolor}
\usepackage[colorlinks=true,allcolors=blue]{hyperref}

\graphicspath{{figs/}}

\newcommand{\cdag}{c^{\dagger}}
\newcommand{\Meff}{\mathcal{M}}
\newcommand{\Tr}{\mathrm{Tr}}
\newcommand{\avg}[1]{\langle #1 \rangle}

\begin{document}

\title{Number Fluctuations and Entanglement-Spectrum Participation in Monitored Free Fermions}

\author{Enso O. Torres Alegre}
\email{etorresa@nd.edu}
\affiliation{Department of Civil and Environmental Engineering and Earth Sciences, University of Notre Dame, Indiana 46556, USA}

\begin{abstract}
On finite system sizes, monitored one-dimensional free-fermion chains display a
broad crossover in entanglement scaling as the measurement rate is increased,
from sub-extensive behavior at weak monitoring toward an area law at strong
monitoring. Analytical field theory and recent large-scale simulations indicate
that this apparent change does not constitute a finite-rate transition in the
thermodynamic limit. I test, with trajectory-resolved
correlation-matrix simulations (chains up to $L=96$, up to $128$ Born-rule
trajectories per parameter point, errors estimated across independent
trajectories), whether two quantities built from the single-particle entanglement
spectrum add practical value to this problem: the bipartite particle-number
fluctuation $F_A=\sum_k\nu_k(1-\nu_k)$ and a participation-style ``effective
number of entangling modes'' $\Meff_A=\exp[-\sum_k w_k\ln w_k]$, with
$w_k\propto\nu_k(1-\nu_k)$. These results are mixed, and I report them as such.
(i)~Both quantities track the entanglement crossover: they decrease monotonically
with the measurement probability and lose their system-size dependence in the
strongly monitored regime. (ii)~However, at the level of trajectory- and
time-averaged steady-state values, $\Meff$ is very nearly a deterministic
function of $F$: a single pooled curve captures $99.7\%$ of its variance across
all measurement rates, so $\Meff$ carries little information that is
independent of the number fluctuation for this model. (iii)~The one genuine
practical advantage I can substantiate is statistical: the trajectory-to-trajectory
coefficient of variation of $\Meff$ is up to a factor ${\sim}2$ smaller than that
of the entropy or of $F$ in the strongly monitored regime. (iv)~Model comparison
of the subsystem scaling at $L=48$--$96$ shows that at weak monitoring a pure
logarithmic law is \emph{not} the statistically preferred description at these
sizes: a small residual quasi-extensive component is detected, and the effective
logarithmic coefficient drifts with $L$, consistent with the crossover
(rather than critical-phase) scenario supported by recent field-theory analyses.
I conclude that $F$ remains the natural experimentally motivated companion to
the entropy, while $\Meff$ is best regarded as a numerically convenient,
variance-reduced summary of the same information, not an independent diagnostic.
\end{abstract}

\maketitle

\section{Introduction}
Repeated local measurements interspersed with unitary evolution can qualitatively
reshape the entanglement structure of a many-body state. In generic (interacting)
circuits this gives rise to a measurement-induced phase transition (MIPT)
separating a volume-law entangled phase at weak monitoring from an area-law phase
at strong monitoring~\cite{Li2018,Skinner2019,Chan2019,Li2019}. The transition has
been characterized as a purification transition~\cite{Gullans2020a,Gullans2020b},
mapped onto effective statistical-mechanics models and replica field
theories~\cite{Bao2020,Jian2020,Choi2020}, and probed on noisy quantum
hardware~\cite{Noel2022,Koh2023,Hoke2023}. Overviews of the phenomenology can be found in
recent reviews~\cite{Fisher2023,PotterVasseur2022}.

Free-fermion (Gaussian) chains under continuous or projective monitoring occupy a
special and much-studied corner of this
landscape~\cite{Cao2019,Alberton2021,Buchhold2021,Szyniszewski2019,Turkeshi2021,Coppola2022,Kells2023}.
Because Gaussian states are fully specified by a two-point function, trajectories
can be simulated for large systems without postselection, and the entanglement
entropy is obtained from the eigenvalues of a reduced correlation
matrix~\cite{Peschel2003,Peschel2009}. The emerging picture in one dimension is
subtle. At weak monitoring the entanglement entropy grows sub-extensively and is
well described, over accessible length scales, by an approximately logarithmic
law with a non-universal, measurement-rate--dependent
coefficient~\cite{Alberton2021,Buchhold2021}. At strong monitoring the entropy
saturates to an area law. Whether the weak-monitoring behavior survives as a
distinct phase in the thermodynamic limit has been seriously questioned:
nonlinear sigma-model analyses~\cite{Fava2023} and a renormalization-group
treatment of free fermions under random projective measurements~\cite{Poboiko2023}
argue that in strictly one dimension the logarithmic regime is a long-lived
crossover that ultimately flows to an area law for any nonzero measurement rate.
Recent large-scale GPU simulations reaching $L=16384$ provide direct numerical
support for this scenario for both projective and homodyne monitoring and indicate
that sizes of order $L\sim10^4$ may be required to resolve the asymptotic
one-dimensional behavior~\cite{Fan2026}. Although the projective protocol of
Ref.~\cite{Fan2026} is not identical to the discrete-time implementation used
here, its asymptotic conclusion is directly relevant to the interpretation of
finite-size scaling. The purpose of the present study is therefore not to
determine the thermodynamic phase diagram, but to evaluate the information
content and statistical performance of finite-size diagnostics at computationally
accessible sizes. I take this conclusion seriously throughout, and, as I show
below, my own finite-size analysis is consistent with it.

Against this backdrop, a practical question is which observables most conveniently
track the change in entanglement scaling on the system sizes and trajectory
counts that are actually accessible. R\'enyi entropies are the sharpest
theoretical diagnostics, but on hardware they generally demand post-selection or
randomized-measurement machinery~\cite{Ippoliti2021}. Bipartite particle-number
fluctuations are a well-known proxy for entanglement in free-fermion
systems~\cite{Klich2009,Song2012} and have been used in the monitored
context~\cite{Alberton2021}. Separately, the language of effective dimensions and
participation numbers has proven useful in quantum dynamics, from operator
growth~\cite{Parker2019} to spread complexity of
states~\cite{Balasubramanian2022}.

In this paper I \emph{test}---rather than merely propose---two quantities built
from the single-particle entanglement spectrum $\{\nu_k\}$ of a Gaussian state:
the bipartite number fluctuation $F_A=\sum_k\nu_k(1-\nu_k)$ and a
participation-style ``effective number of entangling modes,''
\begin{equation}
  \Meff_A=\exp\Big[-\sum_k w_k\ln w_k\Big],\qquad
  w_k=\frac{\nu_k(1-\nu_k)}{F_A}.
  \label{eq:Mdef}
\end{equation}
The weight $\nu_k(1-\nu_k)$ is the contribution of mode $k$ to the number
fluctuation and vanishes for empty or filled modes, so $\Meff_A$ counts how many
entanglement modes are appreciably ``half-filled.'' Two points of terminology
and interpretation should be fixed at the outset. First, neither quantity is a
symmetry-resolved entanglement measure in the technical sense of a decomposition
of $\rho_A$ into charge sectors~\cite{Goldstein2018}; $F$ is a fluctuation and
$\Meff$ a participation number, and I refer to them as
\emph{number-fluctuation-based} diagnostics. Second, $\Meff$ measures the
\emph{diversity} of the mode contributions, not their overall magnitude: the
normalization by $F_A$ in Eq.~(\ref{eq:Mdef}) removes the total
amplitude, so a weakly entangled state whose few nonzero weights happen to be
comparable can have a sizable $\Meff$. Any claim about $\Meff$ must therefore be
established empirically, which is what I set out to do.

The conclusions are restrained because the data demand it. Both
quantities track the entanglement crossover, but at the level of trajectory- and
time-averaged steady-state values $\Meff$ turns out to be very nearly a
deterministic function of $F$ (a pooled fit captures $99.7\%$ of its variance
across all measurement rates and $128$ trajectories per point), so it adds little
\emph{information} beyond the number fluctuation in this model. What survives is
a \emph{statistical} advantage: $\Meff$ fluctuates markedly less from trajectory
to trajectory than either $S$ or $F$, by up to a factor ${\sim}2$ in coefficient
of variation at strong monitoring. I also quantify, via explicit model
comparison across $L=48$--$96$, the extent to which the weak-monitoring regime
deviates from a pure logarithmic law at accessible sizes. I regard the negative
part of these findings as useful: it delimits what participation-style summaries
of the entanglement spectrum can and cannot be expected to deliver.

\section{Model and methods}
\subsection{Monitored tight-binding chain}
I consider spinless fermions on a ring of $L$ sites with nearest-neighbor
hopping,
\begin{equation}
  H = -J\sum_{i=1}^{L}\left(\cdag_i c_{i+1} + \cdag_{i+1} c_i\right),
  \qquad c_{L+1}\equiv c_1,
\label{eq:H}
\end{equation}
at half filling, with $J=1$, in discrete time. Each step consists of (i) a
unitary hop generated by Eq.~(\ref{eq:H}) over unit time, followed by (ii) an
independent projective measurement of the local occupation $n_l=\cdag_l c_l$ on
each site with probability $p$. In a given step the set of measured sites is
drawn first; the corresponding outcomes are then sampled sequentially in
ascending site order with Born probabilities computed from the current state.
Since occupation projectors on different sites commute, the joint outcome
distribution is independent of this ordering. Outcomes are never postselected;
every trajectory is kept.

I restrict attention to $p>0$. The point $p=0$ is a singular limit: unitary
evolution from this initial state produces a quench steady state with extensive
entanglement, and the limits $p\to0$, $L\to\infty$, $t\to\infty$ do not commute
\cite{Poboiko2023}. Including $p=0$ on the same axis as the monitored data
invites a misreading of the crossover, so I exclude it from all scaling figures
and treat it only as a reference point in the text.

\subsection{Gaussian formalism}
The initial state is the N\'eel product state $|1010\cdots\rangle$; hopping and
occupation measurements preserve Gaussianity. A pure Gaussian state is fully
specified by the correlation matrix $C_{ij}=\avg{\cdag_i c_j}$, a Hermitian
projector ($C^2=C$). The unitary step acts as $C\to\overline{U}\,C\,U^{\!\top}$
with $U=e^{-ih}$, $h$ being the single-particle Hamiltonian matrix. A projective
measurement of $n_m$ updates $C$ as
\begin{align}
  n_m=1\ (\text{prob. } C_{mm}):\quad
  &C_{ij}\to C_{ij}-\frac{C_{im}C_{mj}}{C_{mm}}, \label{eq:up1}\\
  n_m=0\ (\text{prob. } 1-C_{mm}):\quad
  &C_{ij}\to C_{ij}+\frac{C_{im}C_{mj}}{1-C_{mm}}, \label{eq:up0}
\end{align}
for $i,j\neq m$, after which row and column $m$ are zeroed and $C_{mm}$ is pinned
to the outcome. These rules are re-derived and checked on a two-site example in
Appendix~\ref{app:update}.

Numerical safeguards: after every time step $C$ is re-Hermitized,
$C\to(C+C^\dagger)/2$; eigenvalues of reduced blocks are clipped to
$[10^{-12},1-10^{-12}]$ before entering logarithms; and the purity of the global
$C$ (eigenvalues pinned to $0$ and $1$) is verified to hold at machine precision
throughout the runs, providing a stringent internal check. Outcome probabilities
$C_{mm}$ are clipped to $[0,1]$ before sampling; values outside this range never
exceeded rounding error. All random streams use fixed, logged seeds, and the
complete simulation and analysis code is provided with the manuscript.

\subsection{Observables}
For a contiguous block $A$ of $\ell$ sites, diagonalizing $C_A$ yields the
single-particle entanglement spectrum $\{\nu_k\}\subset[0,1]$, from which I
compute the von Neumann entropy $S_A=-\sum_k[\nu_k\ln\nu_k+(1-\nu_k)\ln(1-\nu_k)]$
\cite{Peschel2003}, the number fluctuation $F_A=\sum_k\nu_k(1-\nu_k)
=\mathrm{Var}(N_A)$, and the participation number $\Meff_A$ of
Eq.~(\ref{eq:Mdef}). One computational remark: once $C_A$ is diagonalized, all
three observables cost the same, so neither $F$ nor $\Meff$ is cheaper than $S$
\emph{in this implementation}; the dominant cost is the diagonalization itself.
$F_A$ alone can also be obtained without diagonalization, as
$F_A=\Tr C_A-\Tr C_A^2$, and---unlike $S$ and $\Meff$---is in principle
accessible experimentally from particle-number statistics
alone~\cite{Klich2009,Song2012}. $\Meff$ requires the individual $\nu_k$ and I
know of no measurement protocol that yields it from number statistics; I
therefore regard it as a numerical diagnostic.

\subsection{Statistical protocol}
For each parameter point I evolve $N_{\rm traj}$ independent trajectories,
discard a transient, time-average each observable over the steady-state window
within each trajectory, and only then average over trajectories. Because
successive time steps within a trajectory are correlated, the time average serves
purely as variance reduction; \emph{all quoted uncertainties are standard errors
over independent trajectories}. The integrated autocorrelation time of
$S_{L/2}(t)$ in the steady state, measured over eight long ($400$-step)
trajectories at $L=48$, is $\tau_{\rm int}=6.1\pm0.7$ steps at $p=0.1$ and
$3.7\pm0.3$ steps at $p=0.4$, so a $45$-step averaging window contains roughly
$7$--$12$ effectively independent samples.

Measurement-rate sweeps use $L\in\{24,32,48\}$ with $T=90$ steps, a transient of
$45$ steps, and $N_{\rm traj}=64$ ($L=24,32$) or $128$ ($L=48$). Subsystem-scaling
runs use $L\in\{48,64,96\}$ with $T=110$--$200$ steps scaled with $L$, transients
of $T/2$, and $N_{\rm traj}=24$. As a convergence check, doubling both the
transient and the averaging window at $L=48$ changes the mean half-chain entropy
from $5.950\pm0.029$ to $5.918\pm0.020$ at $p=0.1$ and from $1.925\pm0.018$ to
$1.892\pm0.013$ at $p=0.4$, i.e.\ shifts of at most $1.6$ combined standard
errors; I regard the base windows as adequate for the accuracy quoted here.
Blocks in subsystem scans start at site $1$ of the ring; by translation
invariance of the ensemble (restored after trajectory averaging) the choice of
origin is immaterial, and I verified on a subset of runs that averaging over
block positions is consistent within errors.

\subsection{Code and data availability}

The simulation code, analysis scripts, summary data, and trajectory-resolved
data used to generate the figures are available in a public GitHub repository:
\url{https://github.com/Enso-bio/monitored-free-fermion-diagnostics}.

\section{Results}
\subsection{Entanglement across the monitoring axis}
\begin{figure}[t]
  \centering
  \includegraphics[width=\columnwidth]{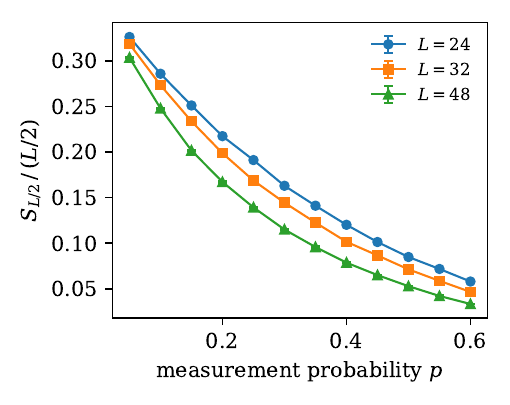}
  \caption{Steady-state half-chain entanglement entropy density versus
  measurement probability $p$ for $L=24,32,48$. Error bars are standard errors
  over $64$--$128$ independent trajectories (often smaller than the symbols).
  The singular unitary point $p=0$ is excluded (see text). I do not interpret
  any particular $p$ as a critical point.}
  \label{fig:scaling_p}
\end{figure}

Figure~\ref{fig:scaling_p} shows the steady-state half-chain entropy density
$S_{L/2}/(L/2)$ versus $p\in[0.05,0.6]$ for three sizes, with standard errors
over trajectories. The entropy is suppressed monotonically by monitoring and
becomes weakly size dependent for $p\gtrsim0.5$. For reference, at $p=0$
(unitary quench, excluded from the figure for the reasons given above) the
half-chain entropy at $L=48$ is $S_{L/2}\approx9.4$, consistent with the
extensive quench value. The curves at different $L$ do not collapse in the
weak-monitoring regime because the entanglement there is sub-extensive rather
than volume law; the progressive loss of size dependence with increasing $p$ is
the expected fingerprint of the crossover to the area law.

\subsection{Subsystem scaling and model comparison}
\begin{figure*}[t]
  \centering
  \includegraphics[width=0.95\textwidth]{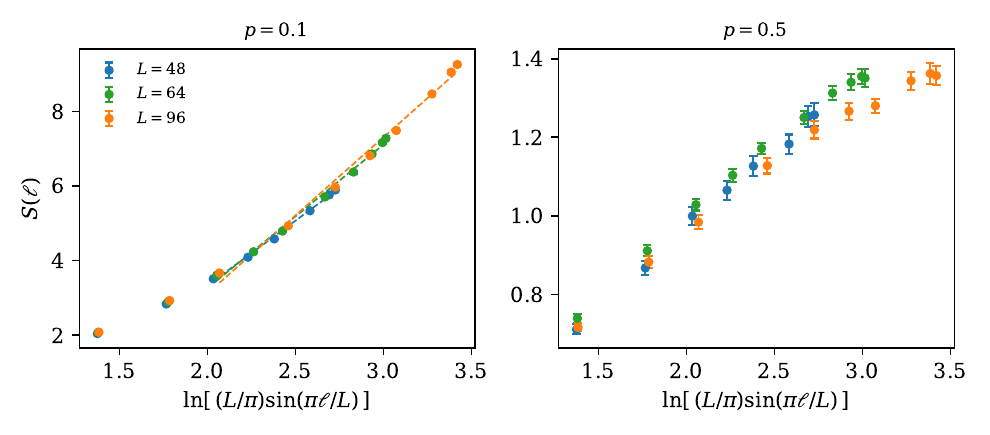}
  \caption{Block entropy versus the logarithm of the conformal chord length for
  $L=48,64,96$ (error bars: standard errors over $24$ trajectories). Left,
  $p=0.10$: the sizes collapse in the chord variable and the data are
  approximately logarithmic (dashed: pure-log fits for $\ell\ge8$), but model
  comparison detects a small residual quasi-extensive component and the slope
  drifts with $L$ (Table~\ref{tab:fits}). Right, $p=0.50$: saturation to an
  area law.}
  \label{fig:subsystem}
\end{figure*}

Figure~\ref{fig:subsystem} shows the block entropy $S(\ell)$ against the
logarithm of the conformal chord length $(L/\pi)\sin(\pi\ell/L)$ for
$L=48,64,96$ at weak ($p=0.10$) and strong ($p=0.50$) monitoring. At $p=0.10$
the data for the three sizes collapse onto a common curve in the chord variable
and are close to linear, i.e.\ approximately logarithmic---formally the
conformal form of Ref.~\cite{Calabrese2004}---while at $p=0.50$ the entropy
saturates, signaling the area law.

``Close to linear,'' however, is not a statistical statement, so I compare two
models on the blocks with $\ell\ge8$:
\begin{align}
  \text{(log)}\quad   &S=\alpha\ln x+b, \nonumber\\
  \text{(log+lin)}\quad &S=\alpha\ln x+\beta\,\ell+b,
  \qquad x=\tfrac{L}{\pi}\sin\tfrac{\pi\ell}{L}, \nonumber
\end{align}
using least squares with parameter uncertainties from the fit covariance and the
Akaike information criterion (AIC) for model selection.
Table~\ref{tab:fits} summarizes the outcome. At $p=0.10$ the log+lin model is
preferred at $L=64$ and $96$ ($\Delta\mathrm{AIC}\simeq-11$ to $-12$), with a
small but statistically significant quasi-extensive coefficient
$\beta\approx0.03$; moreover, the pure-log slope $\alpha$ drifts upward with
system size ($3.48\pm0.09\to4.19\pm0.15$ from $L=48$ to $96$). Both facts say
the same thing: at these sizes and this measurement rate, a pure logarithm is
\emph{not} yet the correct asymptotic description---a residual quasi-extensive
component, inherited from the vicinity of the singular $p=0$ limit, is still
present, and part of the large apparent $\alpha$ absorbs it. At $p=0.20$ the
extensive component is three to five times smaller and $\alpha$ is stable in
$L$ within errors ($\approx2.0$--$2.1$), so the logarithmic form is a good
effective description there. At $p=0.50$ the fitted slopes are tiny
($\alpha\lesssim0.4$) and the entropy is flat at large $\ell$: area law. I
emphasize that even where the logarithmic form fits well, $\alpha$ is an
effective, monitoring-dependent coefficient---not a central charge---and that
the size drift at weak monitoring is precisely what the crossover scenario of
Refs.~\cite{Poboiko2023,Fava2023} leads one to expect.

\begin{table}[t]
\caption{Model comparison for the subsystem scaling, blocks $\ell\ge8$.
$\alpha_{\rm log}$: slope of the pure-log fit; $\beta$: quasi-extensive
coefficient of the log+lin fit; $\Delta\mathrm{AIC}=\mathrm{AIC}_{\rm
log+lin}-\mathrm{AIC}_{\rm log}$ (negative favors log+lin). Parentheses:
$1\sigma$ uncertainties on the last digits.}
\label{tab:fits}
\begin{ruledtabular}
\begin{tabular}{ccccc}
$p$ & $L$ & $\alpha_{\rm log}$ & $\beta$ & $\Delta$AIC \\
\hline
0.10 & 48 & 3.48(9)  & 0.021(10) & $-3.7$ \\
0.10 & 64 & 3.85(12) & 0.033(7)  & $-12.2$ \\
0.10 & 96 & 4.19(15) & 0.029(6)  & $-11.1$ \\
0.20 & 48 & 1.97(3)  & 0.006(3)  & $-2.8$ \\
0.20 & 64 & 2.07(4)  & 0.011(3)  & $-10.5$ \\
0.20 & 96 & 2.11(3)  & 0.006(2)  & $-7.9$ \\
0.50 & 48 & 0.37(2)  & 0.002(3)  & $+0.5$ \\
0.50 & 64 & 0.34(1)  & $-0.003(1)$ & $-15.1$ \\
0.50 & 96 & 0.27(2)  & $-0.003(1)$ & $-6.4$ \\
\end{tabular}
\end{ruledtabular}
\end{table}

\subsection{Number-fluctuation diagnostics track the crossover}
\begin{figure*}[t]
  \centering
  \includegraphics[width=\textwidth]{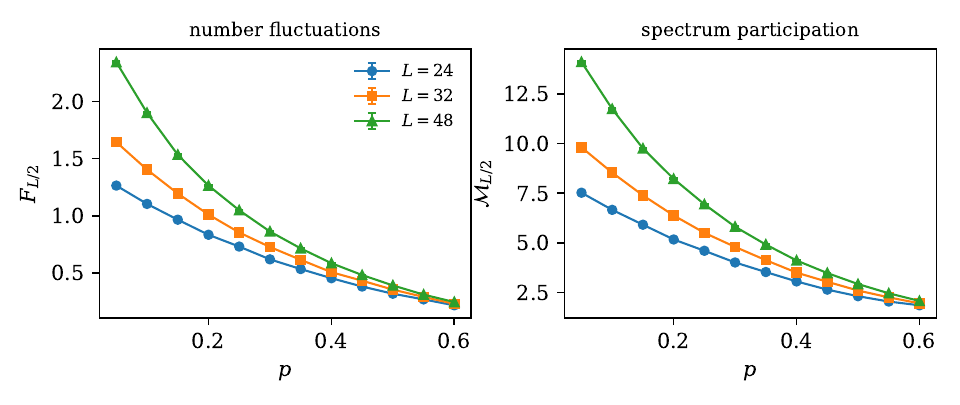}
  \caption{Number-fluctuation-based diagnostics versus measurement probability
  for $L=24,32,48$, with standard errors over trajectories. Left: bipartite
  number fluctuation $F_{L/2}$. Right: spectrum-participation number
  $\Meff_{L/2}$ [Eq.~(\ref{eq:Mdef})]. Both mirror the entropy crossover of
  Fig.~\ref{fig:scaling_p}. No critical point is extracted.}
  \label{fig:diagnostic}
\end{figure*}

Figure~\ref{fig:diagnostic} shows $F_{L/2}$ and $\Meff_{L/2}$ versus $p$ for
$L=24,32,48$ with trajectory standard errors. Both decrease monotonically with
$p$, are clearly separated by system size at weak monitoring, and merge onto a
size-independent curve at strong monitoring---the same qualitative signature as
the entropy. This confirms that both quantities \emph{track} the crossover. The
two questions that matter, however, are whether they add information and whether
they offer statistical advantages; I address them in turn.

\subsection{Does $\Meff$ carry information beyond $F$?}
\label{sec:scatter}
\begin{figure}[t]
  \centering
  \includegraphics[width=\columnwidth]{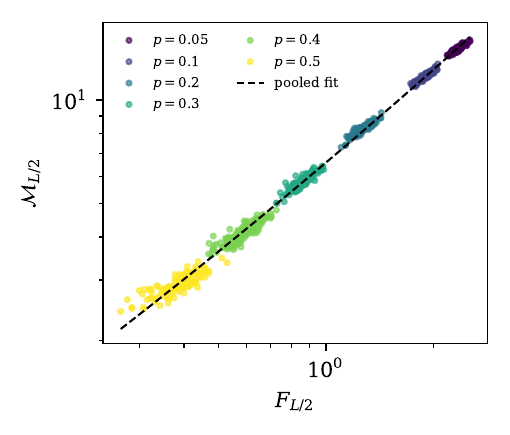}
  \caption{Per-trajectory steady-state averages of $\Meff_{L/2}$ versus
  $F_{L/2}$ at $L=48$ for six measurement rates ($128$ trajectories each,
  log--log scale). The dashed line is a single quadratic fit in the log
  variables pooled over all rates ($R^2=0.997$). The near-collapse means
  $\Meff$ carries little information beyond $F$ in this model; residual
  rate-dependent deviations are $\lesssim1.5\%$ (see text).}
  \label{fig:scatter}
\end{figure}

Figure~\ref{fig:scatter} answers the information question directly. Each point
is one trajectory's steady-state time average $(F_{L/2},\Meff_{L/2})$ at
$L=48$; colors label measurement rates from $p=0.05$ to $0.5$ ($128$
trajectories each). The points fall, to a remarkable degree, on a single smooth
curve: a pooled quadratic fit in $\ln\Meff$ versus $\ln F$ captures
$R^2=0.997$ of the variance across all rates. Deviations from the pooled curve
show small systematic dependence on $p$: group-mean residuals reach at most
$0.015$ in $\ln\Meff$ (i.e.\ ${\sim}1.5\%$ in $\Meff$), which is statistically
resolved at these trajectory counts (up to ${\sim}9$ standard errors at $p=0.2$)
but physically marginal. Within a fixed $p$, the trajectory-to-trajectory
fluctuations of $F$ and $\Meff$ are also strongly correlated (Pearson
$r=0.84$--$0.94$).

The plain-language conclusion is a negative result, and I state it as such: at
the level of trajectory- and time-averaged steady-state values in this model,
$\Meff$ is essentially a monotonic reparametrization of $F$, and knowing
$\Meff$ in addition to $F$ adds almost nothing. Whatever independent structure
of the entanglement spectrum $\Meff$ is sensitive to in principle, the monitored
free-fermion steady state does not exercise it: spectra with equal $F$ but
substantially different participation, which would separate the curves, do not
occur with appreciable probability. Whether models with a genuine transition
(interacting circuits, higher dimensions) populate that region is an open
question I do not answer here.

\subsection{Statistical performance}
\label{sec:stats}
\begin{figure}[t]
  \centering
  \includegraphics[width=\columnwidth]{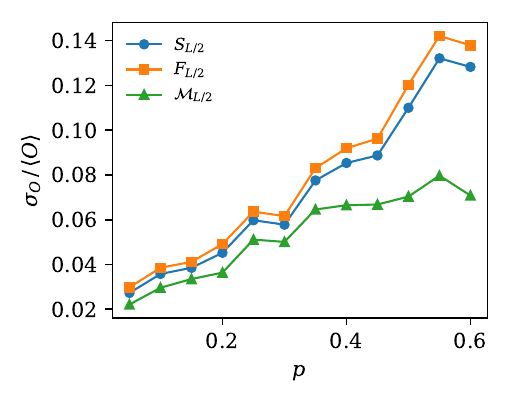}
  \caption{Trajectory-to-trajectory coefficient of variation of the
  steady-state $S_{L/2}$, $F_{L/2}$ and $\Meff_{L/2}$ at $L=48$ ($128$
  trajectories per point). $F$ is marginally noisier than $S$ throughout;
  $\Meff$ is quieter, by up to a factor ${\sim}2$ at strong monitoring.}
  \label{fig:statistics}
\end{figure}

Figure~\ref{fig:statistics} addresses the efficiency question with the
trajectory-to-trajectory coefficient of variation (CV),
$\sigma_O/\avg{O}$, of the three observables at $L=48$ ($128$ trajectories per
point). Two facts emerge. First, contrary to what one might hope, $F$ is
\emph{not} statistically quieter than the entropy: its CV slightly exceeds that
of $S$ at every measurement rate. Second, $\Meff$ \emph{is} quieter, modestly at
weak monitoring and increasingly so at strong monitoring, where its CV
(${\approx}0.07$) is roughly half that of $S$ or $F$ (${\approx}0.13$--$0.14$).
The mechanism is transparent from Eq.~(\ref{eq:Mdef}): the normalization by
$F_A$ cancels the overall amplitude fluctuations that dominate the variance of
$S$ and $F$ deep in the area-law regime, leaving only the (smaller) shape
fluctuations of the spectrum. The same normalization is what makes $\Meff$
nearly redundant with $F$ on average (Sec.~\ref{sec:scatter}); one cannot have
one without the other. I did not observe numerical instability from the
$F_A\to0$ normalization at the rates studied ($F_{L/2}\gtrsim0.2$ at
$p\le0.6$), but $\Meff$ is undefined in the strict $F_A\to0$ limit and should
be used with care very deep in the area-law phase. In terms of fixed-budget
accuracy, the factor ${\sim}2$ in CV translates into a factor ${\sim}4$ fewer
trajectories for equal relative error on $\Meff$ than on $S$ at strong
monitoring---a modest but genuine practical advantage for finite-size studies
that need a smooth scalar tracking the crossover.

\subsection{Entanglement growth}
\begin{figure}[t]
  \centering
  \includegraphics[width=\columnwidth]{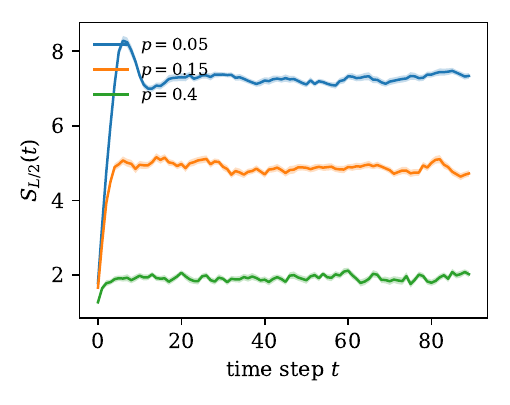}
  \caption{Half-chain entropy versus time from the N\'eel state at $L=48$;
  shaded bands are standard errors over $40$ trajectories. The transient
  overshoot at $p=0.05$ is the coherent quench entanglement peak, subsequently
  degraded by monitoring toward the steady state (see text).}
  \label{fig:growth}
\end{figure}

Figure~\ref{fig:growth} shows the build-up of $S_{L/2}(t)$ from the N\'eel
state at $L=48$ with standard-error bands. Stronger monitoring suppresses the
saturation value, and the initial rise is visibly truncated earlier at larger
$p$; I refrain from quantitative statements about the initial growth
\emph{rate}, which would require a dedicated fit of the early-time window. The
$p=0.05$ curve displays a clear transient overshoot: it peaks near
$S\approx8.3$ around $t\approx20$ before relaxing to a plateau near $7.1$. The
peak position matches the ballistic time $t\sim(L/2)/v_{\max}$ (with
$v_{\max}=2J$) at which counter-propagating quasiparticle pairs from the quench
have maximally entangled the two halves of the ring~\cite{Nahum2017}; under
weak monitoring this coherent quench entanglement is subsequently degraded
toward the lower monitored steady-state value, producing the overshoot. It is a
finite-size, weak-monitoring transient---absent at $p=0.15$ and $0.40$---and
the steady-state windows used elsewhere in the paper start well after it has
decayed (Sec.~II.D).

\section{Discussion}
It is worth being precise about what this study establishes.

\emph{Established.} (i) Both number-fluctuation-based quantities track the
monitored free-fermion entanglement crossover on accessible sizes, with the
expected loss of size dependence at strong monitoring. (ii) The steady-state
participation number $\Meff$ is, to within ${\sim}1.5\%$, a function of the
number fluctuation $F$ across the entire range of measurement rates studied:
it does not constitute an independent diagnostic in this model. (iii) $\Meff$
does offer a quantifiable statistical advantage (up to ${\sim}2\times$ smaller
CV, i.e.\ ${\sim}4\times$ fewer trajectories at fixed relative error, at strong
monitoring), while $F$ offers none over $S$. (iv) At $p=0.1$ and sizes up to
$L=96$, a pure logarithmic subsystem scaling is statistically disfavored
against log plus a small extensive component, and the effective log coefficient
drifts with $L$; the weak-monitoring regime at these sizes retains memory of
the extensive $p=0$ limit, consistent with the crossover picture of
Refs.~\cite{Poboiko2023,Fava2023,Fan2026}.

\emph{Not established, deliberately.} I do not locate a critical measurement
rate, and I do not claim one exists in this model; these data are consistent
with a broad crossover. My maximum size, $L=96$, is far below the scale at which
the asymptotic one-dimensional behavior becomes numerically visible in
Ref.~\cite{Fan2026}; the scaling analysis should therefore be read as a
finite-size diagnostic study rather than an independent determination of the
thermodynamic phase diagram. I do not claim $\Meff$ reveals structure of the
entanglement spectrum invisible to simpler observables---my own scatter
analysis shows the opposite here. I do not claim experimental accessibility
for $\Meff$: unlike $F$, which is a particle-number statistic~\cite{Klich2009,
Song2012}, $\Meff$ requires the individual spectrum eigenvalues, for which I
know no measurement protocol. And the ``cheaper'' framing sometimes attached to
spectrum-derived diagnostics does not apply: all three observables share the
cost of diagonalizing $C_A$; only $F$ escapes it via $F_A=\Tr C_A-\Tr C_A^2$.

\emph{Scope.} The Gaussian restriction is essential to this method and excludes
interactions, which change the universality class of the problem and stabilize
a genuine volume-law phase. The redundancy of $\Meff$ with $F$ found here is a
statement about the monitored free-fermion steady-state ensemble, not a
mathematical identity; models whose entanglement spectra explore shapes at
fixed $F$---interacting monitored circuits, higher-dimensional fermions,
symmetry-enriched settings---could in principle separate the two, and the
scatter analysis of Sec.~\ref{sec:scatter} is a cheap, transferable test for
exactly that. Whether $\Meff$ acquires genuine diagnostic power in a model with
a true transition is, in my view, the right next question, and the honest
prior after this study is skeptical.

\section{Conclusion}
I aimed to test whether a participation-style summary of the single-particle
entanglement spectrum adds practical value to the study of monitored free
fermions, alongside the familiar bipartite number fluctuation. The answer I
can defend is limited and partly negative: both quantities track the
entanglement crossover, but the participation number is informationally almost
redundant with the number fluctuation in this model ($R^2=0.997$ pooled across
measurement rates), and its one demonstrated advantage is a factor ${\sim}2$
reduction in trajectory-to-trajectory noise at strong monitoring. This model
comparison also quantifies how far the weak-monitoring regime is from a clean
logarithmic law at sizes up to $L=96$, in line with analytical and large-scale
numerical evidence for the crossover interpretation of the one-dimensional
problem~\cite{Poboiko2023,Fava2023,Fan2026}. I believe negative results of
this kind---delimiting what cheap spectral summaries can deliver---are useful
to a field in which candidate diagnostics multiply quickly, and I release the
full trajectory-resolved dataset and code so that the same tests can be applied
to models where the outcome may differ.

\begin{acknowledgments}
The author acknowledges the use of OpenAI Codex for assistance with simulation
code development and debugging, and ChatGPT-5.5 for language editing,
consistency checks, and manuscript revision support. All AI-assisted outputs
were reviewed and verified by the author, who takes full responsibility for the
scientific content, calculations, code, figures, and conclusions. This work was
completed prior to the author joining the University of Notre Dame.

\end{acknowledgments}

\appendix
\section{Gaussian measurement update}
\label{app:update}
Conditioning a pure Gaussian state on the outcome of a local occupation
measurement yields another pure Gaussian state whose correlation matrix follows
from a Schur-complement update, Eqs.~(\ref{eq:up1})--(\ref{eq:up0}). As a
minimal check, take two sites in the one-particle state
$|\psi\rangle=\tfrac{1}{\sqrt2}(\cdag_1+\cdag_2)|0\rangle$, with
\begin{equation}
  C=\begin{pmatrix} 1/2 & 1/2\\ 1/2 & 1/2\end{pmatrix},
\end{equation}
a rank-one projector. Measuring $n_1$ gives the occupied outcome with
probability $C_{11}=1/2$; Eq.~(\ref{eq:up1}) then yields
$C_{22}\to\tfrac12-\tfrac{(1/2)(1/2)}{1/2}=0$ and, after pinning,
$C=\mathrm{diag}(1,0)$, i.e.\ the state $\cdag_1|0\rangle$. The empty outcome
gives, via Eq.~(\ref{eq:up0}), $C_{22}\to\tfrac12+\tfrac{(1/2)(1/2)}{1/2}=1$,
i.e.\ $C=\mathrm{diag}(0,1)$, the state $\cdag_2|0\rangle$. Both
post-measurement matrices are projectors, confirming that the update preserves
Gaussian purity; in the many-site simulations this purity is maintained to
machine precision.


\end{document}